# Selective electrochemical generation of hydrogen peroxide from water oxidation

Venkatasubramanian Viswanathan,[*,†,§] Heine A. Hansen,[‡,§] and Jens K. Nørskov[*,¶,∥]

*Department of Mechanical Engineering, Carnegie Mellon University, Pittsburgh, PA 15213, USA, Department of Energy Conversion and Storage, Technical University of Denmark, Kgs. Lyngby DK-2800, Denmark, and SUNCAT Center for Interface Science and Catalysis, Department of Chemical Engineering, Stanford University, Stanford, California, 94305-3030, USA*

E-mail: venkvis@cmu.edu; norskov@stanford.edu

Water is a life-giving source, fundamental to human existence, yet, over a billion people lack access to clean drinking water. Present techniques for water treatment such as piped, treated water rely on time and resource intensive centralized solutions. In this work, we propose a decentralized device concept that can utilize sunlight to split water into hydrogen and hydrogen peroxide. The hydrogen peroxide can oxidize organics while the hydrogen bubbles out. In enabling this device, we require an electrocatalyst that can oxidize water while suppressing the thermodynamically favored oxygen evolution and form hydrogen peroxide. Using density functional theory calculations, we show that the free energy of adsorbed OH* can be used as a descriptor to screen for selectivity trends between the 2e$^-$ water oxidation to $H_2O_2$ and the 4e$^-$ oxidation to $O_2$. We show that materials that bind oxygen intermediates sufficiently weakly, such as $SnO_2$, can activate hydrogen peroxide evolution. We present a rational design principle for the selectivity in electrochemical water oxidation and identify new material candidates that could perform $H_2O_2$ evolution selectively.

[*]To whom correspondence should be addressed
[†]Department of Mechanical Engineering, Carnegie Mellon University, Pittsburgh, PA 15213, USA
[‡]Department of Energy Conversion and Storage, Technical University of Denmark, Kgs. Lyngby DK-2800, Denmark
[¶]SUNCAT Center for Interface Science and Catalysis, Department of Chemical Engineering, Stanford University, Stanford, California, 94305-3030, USA
[§]SUNCAT Center for Interface Science and Catalysis, Department of Chemical Engineering, Stanford University, Stanford, California, 94305-3030, USA
[∥]SUNCAT Center for Interface Science and Catalysis, SLAC National Accelerator Laboratory, Menlo Park, CA 94025-7015, USA

The global energy consumption is projected to increase with the increased energy consumption being concentrated in areas that rank highest on the water scarcity index.[1,2] A key challenge in providing clean drinking water is to find a low-cost, energy-efficient approach to treatment and disinfection of water so that it is suitable for consumption. Providing piped, treated water requires time and resource intensive centralized facilities and an infrastructure that does not exist in many places today.[3] Conventional techniques for water disinfection typically involve the use of chlorine or ozone as the oxidant.[4] However, chlorination generates hazardous and carcinogenic compounds,[5] while the use of ozone, though efficient and harmless, is expensive.[4] Hydrogen peroxide is an attractive candidate for water treatment as its degradation product is water and it has strong oxidative properties.[6]

$H_2O_2$ is generated at an industrial scale through the oxidation of anthraquinone. This process is a multi-step method requiring significant energy input and generates substantial waste and its transportation causes possible hazards.[6] A direct efficient and economic route for production of hydrogen peroxide could enable its wide-spread application in water treatment and other areas. The direct thermal catalytic synthesis from $H_2$ and $O_2$ on palladium based materials have been studied for many years.[6–8] However, selectivity and production rate of $H_2O_2$ are far below the desired limit.[6] An alternate synthetic route is through direct electrochemical reduction of oxygen and protons.[9,10] Experimental and theoretical studies have shown it is possible to selectively activate hydrogen peroxide generation from oxygen reduction, however, this route requires electrocatalysts that are made out of expensive metals

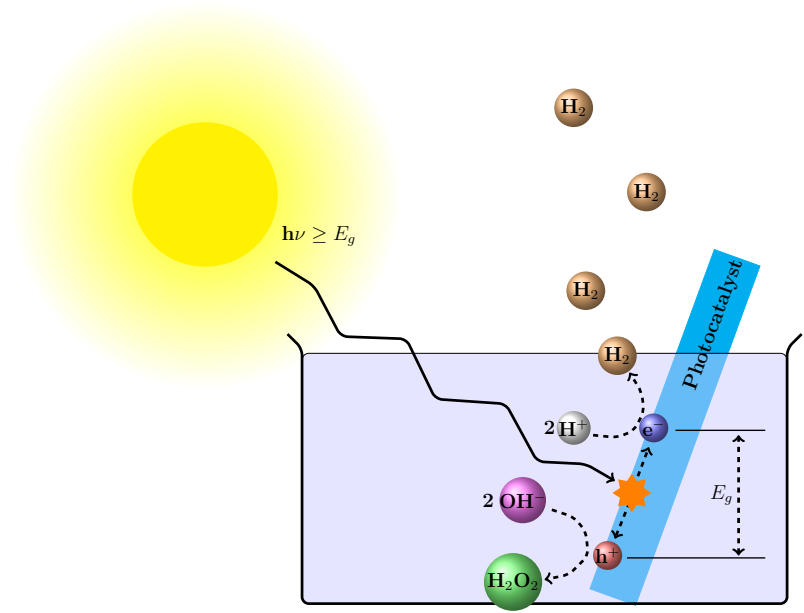

Figure 1: Schematic of a novel 'dream' device concept that can absorb photons from sunlight and use it to split water into hydrogen and hydrogen peroxide. Hydrogen peroxide will decompose the organics in water and thus cleaning water and hydrogen will bubble out.

such as gold, platinum and palladium.[11–14]

Our core idea is to identify a synthetic route for hydrogen peroxide that uses water as its only reactant and uses sustainable electricity preferably from sunlight. In Figure 1, we show the schematic of a decentralized water treatment device. This device employs a material that can absorb photons and generate electrons and holes with appropriate energy such that the electrons can reduce protons to hydrogen and holes can oxidize water to $H_2O_2$. To enable this device, two material challenges need to be overcome. The first is we require a photon absorber whose band positions are suitably aligned such that it can catalyze hydrogen evolution and hydrogen peroxide evolution. The second, more formidable requirement is of an electrocatalyst that can catalyze $H_2O_2$ evolution and suppress the thermodynamically favored $O_2$ evolution. The second electrocatalyst requirement forms the focus of our present work.

In this work, using thermodynamic analysis based on density functional theory calculations, we demonstrate the existence of material candidates that can activate $H_2O_2$ evolution through the oxidation of water. We show that the free energy of adsorbed OH* can be used as a descriptor, to a first approximation, for determining trends in 2e$^-$ vs 4e$^-$ oxidation of H$_2$O. We identify materials that are good candidate materials for H$_2$O$_2$ generation. Among these, we identify SnO$_2$ and TiO$_2$ as candidate materials with good selectivity and our analysis provides a quantitative foundation for the identification of more efficient, selective electrocatalyst materials. This analysis provides a necessary, but not sufficient criterion for a good selective electrocatalyst. Undoubtedly, kinetic barriers are important for determining selectivity, however, such a thermodynamic analysis has proved successful in determining selectivity trends for oxygen reduction.[12,13,15]

## Results

An ideal electrocatalyst for the 4 electron oxygen evolution reaction should be capable of facilitating oxidation of H$_2$O just above the equilibrium potential of 1.23 V. As a minimum requirement, the four charge transfer steps should have reaction free energies of the same magnitude equal to the equilibrium potential of 1.23 eV. We consider the associative mechanism shown below:[16–19]

$$2\text{H}_2\text{O(l)} + * \rightarrow \text{OH}* + \text{H}_2\text{O(l)} + \text{H}^+ + e^-, \tag{1a}$$

$$\text{OH}* + \text{H}_2\text{O(l)} + \text{H}^+ + e^- \rightarrow \text{O}* + \text{H}_2\text{O(l)} + 2\text{H}^+ + 2e^-, \tag{1b}$$

$$\text{O}* + \text{H}_2\text{O(l)} + 2\text{H}^+ + 2e^- \rightarrow \text{OOH}* + 3\text{H}^+ + 3e^-, \tag{1c}$$

$$\text{OOH}* + 3\text{H}^+ + 3e^- \rightarrow \text{O}_2(g) + 4\text{H}^+ + 4e^- + *. \tag{1d}$$

However, an ideal electrocatalyst for the two electron oxidation of water to hydrogen peroxide should facilitate the oxidation just above the equilibrium potential of 1.77 V. This implies that each of the two charge transfer steps must have a reaction free energy of 1.77 eV. We consider a similar associative mechanism for $H_2O_2$ production,

$$2H_2O(l) + * \rightarrow OH* + H_2O(l) + H^+ + e^-, \qquad (2a)$$

$$OH* + H_2O(l) + H^+ + e^- \rightarrow H_2O_2(l) + 2H^+ + 2e^-. \qquad (2b)$$

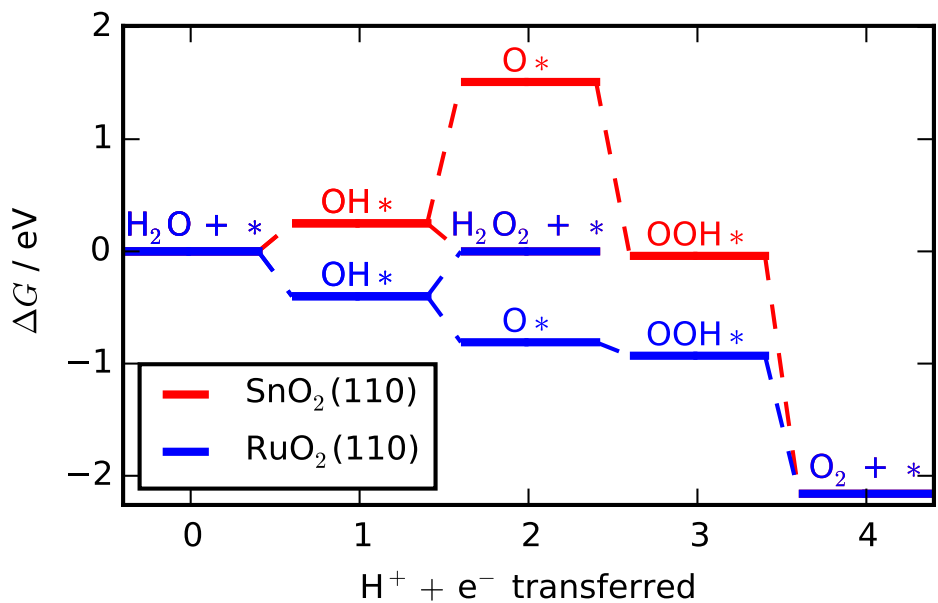

Figure 2: Free energy diagram of water oxidation plotted at U = 1.77 V, versus the reversible hydrogen electrode on $SnO_2(110)$ and $RuO_2(110)$. On $SnO_2(111)$, the limiting step for $2e^-$ oxidation is the activation of $H_2O$ as $OH^*$ while that for the $4e^-$ oxidation is the oxidation of $OH^*$ to $O^*$. On $RuO_2(111)$, the limiting step for $2e^-$ oxidation is the formation of $H_2O_2$ from $OH^*$ while that for the $4e^-$ oxidation is the oxidation of $O^*$ to $OOH^*$.

In Figure 2, we show the calculated free energy diagram for the $2e^-$ and $4e^-$ oxidation of water on rutile type $SnO_2$ and $RuO_2$ at U = 1.77 V, versus the reversible hydrogen electrode.[17] Using the free energy diagrams, an important parameter, the thermodynamic limiting potential, $U_L$, can be extracted and this is defined as the lowest potential at which all of the reaction steps are downhill in free energy. This approach has been successfully used to rationalize trends in hydrogen evolution, oxygen reduction and oxygen evolution.[17,20–22] The potential determining step for the $2e^-$ oxidation on $SnO_2$ is the activation of $H_2O$ as $OH^*$ while that for the $4e^-$ oxidation is the oxidation of $OH^*$ to $O^*$. It is to be noted that the calculated limiting potential for the 2e⁻ oxidation is lower than that for the 4e⁻ oxidation and we would expect this to show selectivity towards hydrogen peroxide generation. In contrast, the potential determining step for 2e⁻ oxidation on $RuO_2$ is the formation of $H_2O_2$ from OH* while the potential determining step for 4e⁻ oxidation is the oxidation of O* to OOH*. For $RuO_2$(110), the activation to OH* is facile at 1.77 V, however, because the further oxidation of OH* to O* is strongly exothermic, selectivity for $H_2O_2$ is expected to be low on $RuO_2$. Materials with an OH* binding energy between that on $SnO_2$ and $RuO_2$ are expected to have improved activity for the 2e⁻ oxidation provided selectivity for the 4e⁻ oxidation can be suppressed.

The trends for oxygen electrochemistry is determined by the binding of three key intermediates, O*, OH* and OOH*.[16] However, it has been demonstrated that the binding of these intermediates on oxide materials are correlated. This enables the activity, given by the limiting potential, $U_L$, to be plotted as a function of a single descriptor, to a first approximation. These plots allow for the quantitative determination of the descriptor values that yield optimal catalyst activity.[16,17,21,23]

Generalizing this analysis, for the 4e⁻ oxidation of water, in the case of materials that bind oxygen intermediates too strongly, we have step 1c associated with the oxidation of adsorbed O* being the limiting step. Therefore, the free energy difference of the limiting step is given by,

$$\Delta G_{1c} = \Delta G_{OOH*} - \Delta G_{O*}. \tag{3}$$

In the case of the materials that bind oxygen intermediates too weakly, we have step 1b associated with the oxidation of adsorbed OH* being the limiting step for the 4e⁻ oxidation of water. Therefore, the free energy difference of the limiting step is given by,

$$\Delta G_{1b} = \Delta G_{O*} - \Delta G_{OH*} \tag{4}$$

For the 2e⁻ oxidation of water, the activity of materials that bind oxygen intermediates too strongly, reaction 2b associated with the oxidation of OH* to $H_2O_2$ is the limiting step. The free energy of the limiting step is given by,

$$\Delta G_{2b} = \Delta G_{H_2O_2(l)} - \Delta G_{OH*}.  \qquad (5)$$

The activity of materials on the weak binding leg of the volcano is limited by 2a associated with the activation of water as OH* and the free energy difference of the limiting step is given by,

$$\Delta G_{2a} = \Delta G_{OH*} - \Delta G_{H_2O(l)} \qquad (6)$$

In this work, we have chosen free energy of OH*, $\Delta G_{OH*}$, as the descriptor and this choice is made as the activity for hydrogen peroxide evolution is directly determined by $\Delta G_{OH*}$. In Figure 3, we show the calculated thermodynamic limiting potentials, $U_L$, as a function of the free energy of OH*, $\Delta G_{OH*}$.

The calculated activity volcano shows that the optimal catalyst for peroxide evolution should exhibit a free energy of OH* adsorption of 1.77 eV. We identify $RuO_2$, $PtO_2$ as materials that bind oxygen too strongly for peroxide evolution. $PtO_2$ calculations are performed on a rutile structure. According to our analysis, we also identify $SnO_2$ and $TiO_2$ as materials that bind oxygen too weakly and have slightly lower potentials for peroxide evolution than oxygen evolution. It is worth highlighting that kinetic barriers are also important for determining selectivity and the activity volcano analysis presented here provides a necessary, but not sufficient criterion for a good selective electrocatalyst.

## Discussion

The only direct experimental demonstration for selectively making $H_2O_2$ over $O_2$ through electrochemical oxidation of water has been the recent work of MacFarlane and co-workers who employed $MnO_x$ electrode with an ionic liquid based electrolyte to tune the thermo-dynamics and showed $H_2O_2$ formation.[25,26] Based on our analysis, we find that the limiting potentials for hydrogen peroxide evolution and oxygen evolution on $MnO_2$ are quite close and small changes in surface energetics due to solvent effects (ionic liquid) could affect the selectivity. There has been no direct experimental demonstration of electrochemical $H_2O_2$ generation in a purely aqueous system. It is worth pointing out that photogeneration of $H_2O_2$ has been demonstrated on $TiO_2$ and ZnO.[27,28]

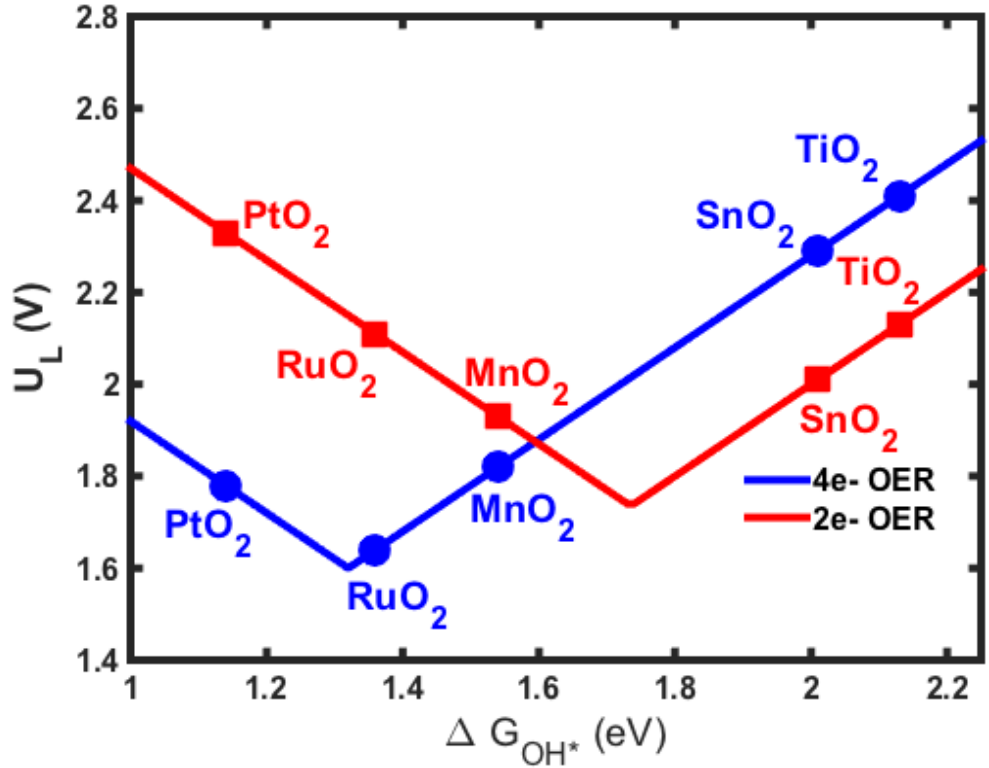

Figure 3: Activity volcano for the 2e$^-$ and 4e$^-$ oxidation of water. The scaling relations used to construct the volcano are $\Delta G_{OOH} = \Delta G_{OH} + 3.2$ [eV] and $\Delta G_O = 2\Delta G_{OH} + 0.28$ [eV].[24]

It has also been demonstrated that electrodes that have high overpotentials for oxygen evolution exhibit enhanced performance towards the decomposition of organics although the exact details of the mechanism are still unclear.[29,30] For instance, it has been shown that $SnO_2$ which is poor at catalyzing oxygen evolution exhibits a higher efficiency for organic degradation compared to Pt.[30] We suggest that generation of hydrogen peroxide is a precursor to the decomposition of organics and based on this assumption, our analysis attributes the enhanced efficiency for the decomposition of organics to the weaker binding of oxygen intermediates on $SnO_2$ compared to Pt. As a result, we would expect little decomposition of organics on $PtO_2$ and significant decomposition on $SnO_2$.

Our descriptor based approach can be used to identify possible candidate materials that could be effective at selectively catalyzing peroxide evolution. An avenue could be the doping of metal ions on cheap material such as $TiO_2$, which has been pursued for oxygen evolution.[31] We have searched for suitable doped $TiO_2$ candidates for selective peroxide evolution. Based on our analysis, we identify $TiO_2$ doped with Ru or Ir are interesting candidate materials that could exhibit enhanced activity for peroxide evolution, as discussed in the SI. In addition to selectivity, stability is a stringent requirement for water oxidation electrocatalysts[32,33] and we expect the stability requirement to be play a crucial role in the selection of an electrocatalyst.

Finally, the provided device requires a photon absorber whose band positions are suitably aligned for hydrogen peroxide generation. This poses a requirement for a photon absorber that the valence band maximum is greater than 1.77 V vs normal hydrogen electode (NHE) and the conduction band minimum is less than 0 V vs NHE. There are many material candidates that satisfy this criterion, including $SnO_2$. Hence, identifying a suitable photon absorber is less challenging than finding a selective electrocatalyst. This analysis suggests that it is possible to identify a single material that can carry out the catalysis as well as the photon absorption.

We have outlined a quantitative framework for determining selectivity in electrochemical water oxidation. We show that it is possible to selectively catalyze the $2e^-$ oxidation to $H_2O_2$ over the thermodynamically favored oxygen evolution. This can be accomplished under certain range of potentials by choosing catalysts that are inefficient at carrying out oxygen evolution. We show $SnO_2$ and $TiO_2$ as materials that exhibit suitable bonding characteristics for peroxide evolution and identify doped $TiO_2$ candidate materials that could carry out this process more efficiently. This shows that it is possible to selectively form fuels or chemicals that involve a smaller number of proton-coupled electron transfer over its thermodynamically favored competing reaction that involves a larger number of proton-electron transfer. We expect this core idea to be broadly useful given the ubiquity of adsorbate scaling relations and we expect it to be particularly useful for nitrogen and carbon electrochemistry.

# Methods

**Free energy diagrams:** The free energy diagram is calculated based on density functional theory calculations which accounts for zero point energy and entropic corrections and a detailed discussion is presented in the SI.[16] The effect of potential, U, is included by shifting the free energy of an electron by -eU and the free energy at a potential U, can thus be calculated using the relation, $\Delta G = \Delta G^0 - eU$ where U is the potential relative to the reversible hydrogen electrode and $\Delta G^0$ is the calculated reaction free energy under standard conditions.[34]

**Computational details:** A detailed description of the computational details is given in the SI.

**Acknowledgements**

The authors acknowledge support from the Department of Energy, Basic Energy Sciences through the SUNCAT Center for Interface Science and Catalysis. V.V. acknowledges helpful discussions with Ramya Yeluri.

**Author contributions:** V.V. and H.A.H conceived the idea and carried out the theoretical calculations. All authors discussed the results and co-wrote the manuscript.

**Additional Information**

**Supplementary Information** accompanies this paper.

**Competing financial interests:** The authors declare no competing financial interests.

# Supporting Information:

# Selective electrochemical generation of hydrogen peroxide from water oxidation

Venkatasubramanian Viswanathan,*,†,§ Heine A. Hansen,‡,§ and Jens K. Nørskov*,¶,∥

*Department of Mechanical Engineering, Carnegie Mellon University, Pittsburgh, PA 15213, USA, Department of Energy Conversion and Storage, Technical University of Denmark, Kgs. Lyngby DK-2800, Denmark, and SUNCAT Center for Interface Science and Catalysis, Department of Chemical Engineering, Stanford University, Stanford, California, 94305-3030, USA*

E-mail: venkvis@cmu.edu; norskov@stanford.edu

---

*To whom correspondence should be addressed
†Department of Mechanical Engineering, Carnegie Mellon University, Pittsburgh, PA 15213, USA
‡Department of Energy Conversion and Storage, Technical University of Denmark, Kgs. Lyngby DK-2800, Denmark
¶SUNCAT Center for Interface Science and Catalysis, Department of Chemical Engineering, Stanford University, Stanford, California, 94305-3030, USA
§SUNCAT Center for Interface Science and Catalysis, Department of Chemical Engineering, Stanford University, Stanford, California, 94305-3030, USA
∥SUNCAT Center for Interface Science and Catalysis, SLAC National Accelerator Laboratory, Menlo Park, CA 94025-7015, USA

# Computational Details

Formation energies of OH∗ and O∗ on $SnO_2$(110) and OH∗, O∗ and OOH∗ $MnO_2$(110) are obtained from Density Functional theory calculations by Man et al.[1] Additional calculations are performed on $SnO_2$(110) to obtain the formation energy of OH∗. It is found that OOH∗ spontaneously transfers a hydrogen atom to a nearby bridging oxygen atom. The free energy of OOH∗ on $SnO_2$(110) is therefore approximated from the scaling relation between OH∗ and OOH∗ and the formation energy of OH∗.[2]

The computational details used are similar to those in ref. 1 and given below for completeness.

Density functional theory calculations are performed with the DACAPO DFT code. Ionic cores are described using Vanderbilt ultrasoft pseudo potentials and Kohn-Sham states are expanded in plane waves with an energy cutoff of 350 eV, while the electron density is expanded in plane waves with an energy cutoff corresponding to 500 eV.[3] Occupation of one-electron states follows a Fermi-Dirac distribution with $k_B T = 0.1\,\text{eV}$ and total energies are extrapolated to $k_B T = 0\,\text{eV}$. Effects of exchange and correlation are described using the RPBE functional.[4]

The $SnO_2$(110) surface is modeled using a slab with a (1 × 2) surface supercell cell consisting of 4 trilayers. The geometry of the bottom two trilayers is fixed in the bulk position, and adsorbates are added to the topside of the slab. Slabs are separated by 15 Å of vacuum and the electrostatic dipole interaction between periodically repeated slabs has been removed.[5] The first Brilluin zone is sampled using 4 × 4 × 1 Monkhorst-Pack grid of k-points.[6] Adsorbates and the two topmost trilayers are optimized until the maximum force component is below 0.05 eV/Å.

# Free Energy Diagrams

Potential dependent free energies are calculated using the computational hydrogen electrode reference.[7] Consider the initial step in water oxidation, written here in acid

$$H_2O(l) + * \longrightarrow OH* + H^+ + e^-. \tag{1}$$

At 0 V vs an reversible hydrogen electrode (RHE) the reaction

$$\tfrac{1}{2}H_2(g) \longrightarrow H^+ + e^- \tag{2}$$

is in equilibrium, so the chemical potential of a proton and an electron is equal to the chemical potential of $\tfrac{1}{2}H_2(g)$

$$\tfrac{1}{2}\mu_{H_2(g)} = \mu_{H^+} + \mu_{e^-}. \tag{3}$$

Therefore the reaction free energy of eq. 1 at 0 V versus RHE, $\Delta G^0$, can be calculated from the equivalent gas phase reaction

$$H_2O(l) + * \longrightarrow OH* + \tfrac{1}{2}H_2(g) \tag{4}$$

$$\Delta G^0 = G(OH*) - G(*) - \mu_{H_2O(l)} + \frac{1}{2}\mu_{H_2(g)}. \tag{5}$$

At an arbitrary potential, $U$ versus RHE, the chemical potential of an electron is shifted by $-eU$ and correspondingly the reaction free energy of eq. 1 is given by

$$\Delta G = \Delta G^0 - eU. \tag{6}$$

The approach is easily generalized to include formation of O* and OOH* adsorbates.

The reaction free energy of eq. 4 is calculated from DFT simulations by adding contributions from entropy and vibrational zero-point energies (ZPE) to the reaction energies obtained from DFT.

$$\Delta G^0 = \Delta E_{DFT} + \Delta E_{\text{ZPE}} - T\Delta S. \qquad (7)$$

For adsorbates, the zero point energy is calculated from vibrational frequencies calculated on $RuO_2(110)$ and taken from ref. 1, while the entropy of adsorbed species is assumed to be negligible. For molecules, the ZPE is obtained from DFT calculated frequencies, while the entropy is taken from experiment.[8] The chemical potential of $H_2O(l)$ is calculated as the chemical potential of $H_2O(g)$ at 0.035 bar, which is the vapor pressure of $H_2O$ at room temperature. The contributions from ZPE and entropy are listed in S1.

Table S1: **Zero point energies and entropic contributions to adsorbates and water oxidation reactions. Energies are in eV.**

|  | $TS$ | $T\Delta S$ | $E_{ZPE}$ | $\Delta E_{ZPE}$ | $\Delta E_{ZPE} - T\Delta S$ |
|---|---|---|---|---|---|
| $H_2O(l)$ | 0.67 |  | 0.56 |  |  |
| $H_2(g)$ | 0.41 |  | 0.27 |  |  |
| $OH*$ | 0.0 |  | 0.36 |  |  |
| $O*$ | 0.0 |  | 0.07 |  |  |
| $OOH*$ | 0.0 |  | 0.43 |  |  |
| $OH* + \frac{1}{2}H_2(g)$ | 0.20 | -0.47 | 0.50 | -0.06 | 0.41 |
| $O* + H_2(g)$ | 0.41 | -0.27 | 0.34 | -0.22 | 0.05 |
| $OOH* + \frac{3}{2}H_2(g)$ | 0.61 | -0.73 | 0.83 | -0.29 | 0.45 |

c

# Screening for new candidate materials

Based on the activity volcano, we can screen for new candidate materials that could posses higher selectivity than $SnO_2$. One strategy is to use a cheap material such as $TiO_2$ and using doping to tune the adsorption energy. Our analysis suggests that a strengthening of the $OH^*$ binding energy by about ∼0.3 eV relative to that of $TiO_2$ could lead to substantially improved selectivity and electrocatalytic activity for $H_2O_2$ evolution.

We screened the calculated adsorption energies on doped rutile $TiO_2(110)$ surfaces based on a substitutional model with 6.25% transition-metal impurities relative to the host Ti atoms in the slab, that is, M-Ti$_{15}$O$_{32}$.[9] The analysis considered transition metals M=V, Nb, Ta, Cr, Mo, W, Mn, Fe, Ru, Ir, and Ni as dopants at four different substitutional sites. Among the considered cases, our analysis suggests the most promising candidates are Ir or Ru-doped TiO$_2$ as shown in Figure S1.

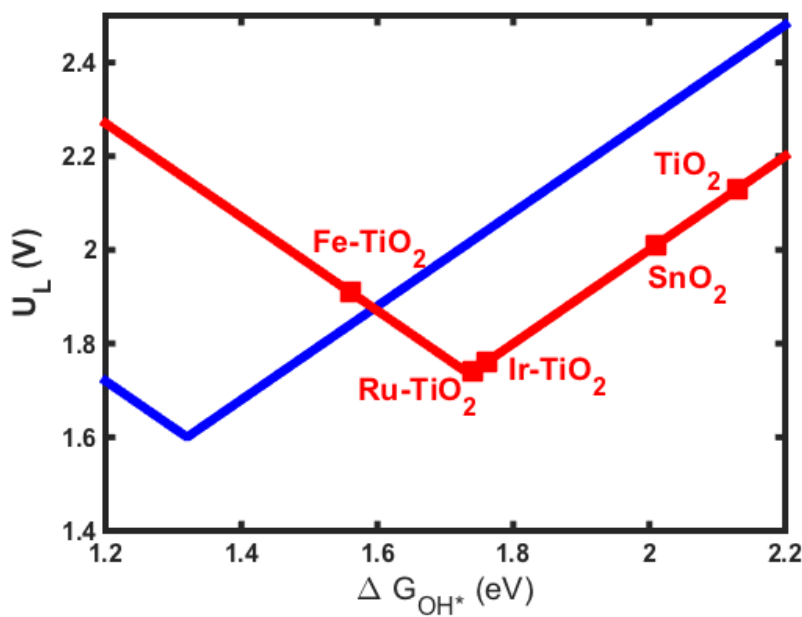

Figure S1: Activity volcano for the 2e$^-$ and 4e$^-$ oxidation of water with the identified doped TiO$_2$ candidates. The adsorption site for the doped TiO$_2$ candidates, Ir and Ru, is on top of a 5-fold coordinated Ti site while the Fe-doped case is on top of a 6-fold coordinated Ti site.